\documentclass{article} % For LaTeX2e
\usepackage{nips15submit_e,times}
\usepackage{hyperref}
\usepackage{url}
\usepackage{soul}
\usepackage{amsmath}
\usepackage{amssymb}  
\usepackage{caption}
\usepackage{subcaption}
\usepackage{graphicx}
\usepackage{wrapfig}

\newcommand{\vb}{{\bf \nu}}

\newcommand{\vx}{{\bf x}}

\newcommand{\vr}{{\bf r}}

\newcommand{\vu}{{\bf u}}

\newcommand{\vl}{{\bf l}}

\newcommand{\vL}{{ L}}
\newcommand{\vK}{{\bf K}}

\newcommand{\fracpartial}[2]{\frac{\partial #1}{\partial  #2}}

\newcommand{\T}{^\mathsf{T}}

\newcommand{\tf}{ t_{f}}
\newcommand{\tfbar}{ \bar{t}_{f}}

\newcommand{\vLx}{\vL_{\vx}}
\newcommand{\vLu}{\vL_{\vu}}
\newcommand{\vLxx}{\vL_{\vx \vx}}
\newcommand{\vLxu}{\vL_{\vx \vu}}
\newcommand{\vLux}{\vL_{\vu \vx}}
\newcommand{\vLuu}{\vL_{\vu \vu}}

\newcommand{\deltaxuo}{\begin{bmatrix} \delta \vx (t) \\ \delta \vu (t)  \end{bmatrix}}
\newcommand{\deltaxuTo}{\bigg[ \delta \vx (t)^{\T}~\delta \vu  (t) ^{\T} \bigg]}
\newcommand{\deltaxbtfo}{\begin{bmatrix} \delta \vx (t) \\ \delta \vb \\ \delta \tf \end{bmatrix}}
\newcommand{\deltaxbtfTo}{\bigg[ \delta \vx (t)^{\T}~\delta \vb^{\T} ~\delta \tf  \bigg]}

\newcommand{\vMatrixVo}{\begin{bmatrix} V_{\vx \vx} &  V_{\vx \vb}  &   V_{\vx \tf}   \\ V_{\vb \vx} &  V_{\vb \vb}  &   V_{\vb \tf} \\ V_{\tf \vx} &  V_{\tf \vb}  &   V_{\tf \tf} \end{bmatrix}}
\newcommand{\vMatrixVxxxo}{\begin{bmatrix} V_{\vx \vx \vx} &  V_{\vx \vx \vb}  &   V_{\vx \vx \tf}   \\ V_{\vx \vb \vx}  &  V_{\vx \vb \vb}   &   V_{\vx \vb \tf}  \\ V_{\vx \tf \vx} &  V_{\vx \tf \vb}   &   V_{\vx \tf \tf}  \end{bmatrix}}
\newcommand{\vMatrixVxxxFo}{\begin{bmatrix} V_{\vx \vx \vx} F &  V_{\vx \vb \vx} F  &   V_{\vx \tf \vx} F   \\ V_{\vb \vx \vx} F &  V_{\vb \vb \vx} F  &   V_{\vb \tf \vx} F \\ V_{\tf \vx \vx} F &  V_{\tf \vb \vx} F  &   V_{\tf \tf \vx} F \end{bmatrix}}
\newcommand{\vMatrixLo}{\begin{bmatrix} \vLxx &  \vLxu \\ \vLux & \vLuu \end{bmatrix}}

\newcommand{\deltaxbtfend}{\begin{bmatrix} \delta \vx + F \delta \tf \\ \delta \vb \\ \delta \tf \end{bmatrix}}

\newcommand{\vMatrixPhiend}{\begin{bmatrix} \Phi_{\vx \vx}  &  \Phi_{\vx \vb} &   \Phi_{\vx \tf}   \\ \Phi_{\vb \vx}  &  \Phi_{\vb \vb}   &   \Phi_{\vb \tf} \\ \Phi_{\tf \vx}  &  \Phi_{\tf \vb}  &   \Phi_{\tf \tf} \end{bmatrix}}

\newcommand{\rd}{\textrm{d}}

\renewcommand{\Re}{\mathbb{R}}

\title{Model Based Reinforcement Learning with Final Time Horizon Optimization}

\author{
Wei Sun \\
School of Aerospace Engineering\\
Georgia Institute of Technology\\
Atlanta, GA 30332-0150, USA. \\
\texttt{wsun42@gatech.edu} \\
\And
Evangelos Theodorou \\
School of Aerospace Engineering\\
Georgia Institute of Technology\\
Atlanta, GA 30332-0150, USA. \\
\texttt{evangelos.theodorou@ae.gatech.edu} \\
\AND
Panagiotis Tsiotras \\
School of Aerospace Engineering\\
Georgia Institute of Technology\\
Atlanta, GA 30332-0150, USA. \\
\texttt{tsiotras@gatech.edu} \\
}

\nipsfinalcopy % Uncomment for camera-ready version

\begin{document}

\maketitle

\begin{abstract}
   We present one of the first algorithms on model based reinforcement learning and trajectory optimization with free final time horizon.  Grounded on the optimal control theory and Dynamic Programming,  we derive a set of   backward differential equations  that propagate the value function and provide the optimal control policy and the optimal time horizon.  The resulting policy generalizes previous results in model based trajectory optimization.   Our analysis   shows  that the  proposed  algorithm  recovers  the theoretical optimal solution on linear  low dimensional problem. Finally we  provide application  results on nonlinear systems. 
\end{abstract}

\section{Introduction}
     
       Trajectory optimization is one of the most active areas of research in machine learning and control theory  with a plethora of applications in robotics,  autonomous systems and computational neuroscience.  
      Among the different methodologies, Differential Dynamic Programming (DDP) is a model based reinforcement learning algorithm that  relies  on linear approximation of dynamics and quadratic approximations of cost functions along nominal trajectories.    
     Even though there has been almost 45 years since the fundamental work by Jacobson and Mayne   on  Differential Dynamic Programming \cite{jacobson1970differential}, it is a fact that research on trajectory optimization and model based reinforcement learning   is performed nowadays  by having a main ingredient DDP.   In the NIPS  community, recently published  state of art methods  on trajectory optimization use DDP   to perform   guided policy search  \cite{NIPS2014_Abbeel} 
      and data-efficient probabilistic trajectory optimization \cite{pan2014probabilistic}.  Earlier work  on DDP  includes  min-max  \cite{morimoto2003minimax},  control limited  \cite{tassa2014control},  receding horizon  \cite{tassa2007receding,abbeel2007application},   and stochastic optimal control formulations \cite{todorov2005generalized,theodorou2010stochastic}. 
     
       Despite all of this research on trajectory optimization using model based reinforcement learning methods such as DDP,   there has not been any effort towards the development of model based trajectory optimization algorithms in which  the  time horizon  is not \textit{a-priori} specified.   The time horizon is one of the important free tuning parameters in  trajectory optimization algorithms and in most case  is manually  tuned  based on the experience of the engineer.        
       
       In this paper we present a new algorithm on model based reinforcement learning in which optimization is performed with respect to control and the time horizon. While  free time horizon DDP has been initially derived by Jacobson and Mayne in \cite{jacobson1970differential},  the resulting algorithm is not implementable and it relies on the assumption that the initialization of the algorithm starts close to the optimal control solution. This will become more clear in the next section as we present our analysis on free final time model based trajectory optimization.

   \section{Problem Formulation and Analysis}
   
   We consider model based reinforcement learning problems in which  optimization occurs with respect to control and time horizon. In mathematical terms these problems are formulated as follows:
  \begin{equation}\label{Cost}
     V(\vx(t_{0}), t_{0}; \vb, \tf)  = \min_{\vu (\cdot) }J(\vx (\cdot) ,\vu (\cdot) )  =  \min_{\vu (\cdot) } \bigg[ \Phi(\vx(\tf), \vb, \tf) +  \int_{t_{0}}^{\tf} {\vL}(\vx (t),\vu (t), t)  \rd t \bigg],
    \end{equation}
where  the term $ \Phi(\vx(\tf), \vb, \tf)  $  is defined  as $\Phi(\vx(\tf), \vb, \tf) = \phi(\vx(\tf), \tf) + \vb^{\T} \psi (\vx(\tf), \tf)$, $\phi(\vx(\tf), \tf)   $  is the terminal cost,  $  \psi (\vx(\tf), \tf)$  is the terminal constraint and $ \vb $ is the corresponding Lagrange mulitplier. ${\vL}(\vx (t),\vu (t), t) $ is the running cost accumulated along the time horizon $ t_{f} $, which is not specified a-priori. The cost function  $J(\vx (\cdot) ,\vu (\cdot) )$ in \eqref{Cost} is minimized subject to the dynamics:
       \begin{equation}
          \frac{\rd \vx(t)}{\rd  t}  = F(\vx(t),\vu(t),t),~~~ \vx_{0} = \vx(t_{0}) .
       \end{equation}
 where $\vx  \in \Re^{n}$  is the state and the $ \vu \in \Re^{m}$  is the control of the dynamics.  Note that the value function $  V(\vx(t_{0}), t_{0}; \vb, \tf)$  is now a function of the Lagrange multiplier  $ \nu $ and the terminal time $ t_{f} $. This is important for the derivation of the free time horizon algorithm since expansions of the value function are computed not only with respect to nominal controls and state trajectories but also with respect to  nominal  $ \bar{\nu} $    and $ \bar{t}_{f} $. 
   
   \subsection{Derivation of Differential Dynamic Programming (DDP) with Free Final Time}
  
        Our analysis and derivation of the free time horizon model based reinforcement learning is in continuous time.  As it is shown, a set of backward ordinary differential equations is derived that back propagates the value function along the nominal trajectory.  In particular,  given a nominal trajectory  $ (\bar{\vx} (\cdot), \bar{\vu} (\cdot))  $ with nominal Lagrange multiplier $\bar{\vb}$ and terminal time $\tfbar$, we start our analysis with the linearization of the dynamics as follows:   
         \begin{align}
      \frac{\rd \vx(t)}{\rd  t} &= F(\bar{\vx} (t) + \delta \vx (t),\bar{\vu} (t) + \delta \vu (t),t) \nonumber \\
       \frac{   \rd      \delta \vx (t) }{\rd t} &=  F_{\vx} (\bar{\vx} (t),\bar{\vu} (t) ,t) \delta \vx (t) +  F_{\vu} (\bar{\vx} (t),\bar{\vu} (t) ,t) \delta \vu (t) .
               \end{align}
All the quantities in the derivation later are evaluated at $(\bar{\vx} (t), \bar{\vu} (t), \bar{\vb}, \tfbar)$ unless otherwise specified. Since our derivation is in continuous time, we consider the  corresponding Hamilton-Jacobi-Bellman equation:
\begin{align} \label{eqn:HJB}
    - \fracpartial{ V(\vx(t), t; \vb, \tf)}{t} =  \min_{\vu(t)} \bigg[ {\cal{H}}(\vx(t), t; \vb, \tf)  \bigg],
\end{align}

under the terminal condition $V(\vx(\tf), \tf; \vb, \tf) = \Phi(\vx(\tf), \vb, \tf)$, and with the Hamiltonian function  $ {\cal{H}}(\vx(t), t; \vb, \tf) $  defined as follows:
\begin{equation}
   {\cal{H}}(\vx(t), t; \vb, \tf) =   {\vL}(\vx(t),\vu(t), t) + V_{\vx} (\vx(t), t; \vb, \tf) ^{\T}  F(\vx(t),\vu(t),t).
\end{equation}
%Before we start our analysis, we would like to point out the difference between our derivation and the derivation in \cite{jacobson1970differential}. 

   We  take expansions of the terms on both sides of \ref{eqn:HJB} around $(\bar{\vx}, \bar{\vu}, \bar{\vb}, \bar{\tf})$. Notice that this is in contrast with the derivation of free final time DDP in \cite{jacobson1970differential} in which the expansion takes place around $(\vx^*, \vu^*)$. Hence, the key assumption in \cite{jacobson1970differential}  is that $\bar{\vu}$ is close to the optimal control $\vu^*$, which makes the algorithm hard to implement, especially when the optimal $\tf$ is not known a-priori. Moreover, expansion of the Hamiltonian $\cal{H}$ around $\vu^*$ yields $\frac{\partial\cal{H}}{\partial \vu} |_{\vu = \vu^*} = 0$, which results in dropping terms from the derivation. 
   
   The left-hand side of (\ref{eqn:HJB}) can be expanded as
                \begin{align}\label{eqn:ExpansionOfPartialV}
                 \fracpartial{}{t} V (\bar{\vx}(t)+ \delta \vx(t) , t ;  \bar{\vb} + \delta \vb , \tfbar + \delta \tf  )   &  \approx   ~ \fracpartial{}{t} \bigg(  V (\bar{\vx}(t) , t ; \bar{\vb} , \tfbar)  + V_{\vx}^{\T}  \delta \vx(t)  + V_{\vb}^{\T}  \delta \vb  +  V_{\tf}  \delta \tf  \bigg)  \nonumber \\
             &  +   \fracpartial{}{t} \bigg( \frac{1}{2} \deltaxbtfTo  \vMatrixVo \deltaxbtfo  \bigg) .
         \end{align}   
Next we make use of the fact that 
    \begin{align}
    \frac{ \rd}{ \rd t} (\cdot) = \fracpartial{}{t} (\cdot) + \fracpartial{}{x} (\cdot)^{\T} F(\bar{\vx}(t),\bar{\vu}(t),t) \Rightarrow -  \fracpartial{}{t} (\cdot) =   -    \frac{ \rd}{ \rd t} (\cdot) +   \fracpartial{}{x} (\cdot)^{\T} F(\bar{\vx}(t),\bar{\vu}(t),t).
    \end{align}
   Based on the equation above  we have that 
       \begin{align} \begin{split}   
     &  -  \fracpartial{}{t}  V (\bar{\vx}(t)+ \delta \vx(t) , t ;  \bar{\vb} + \delta \vb , \tfbar + \delta \tf  )  \\
     = & -\frac{ \rd}{ \rd t} \bigg(  V (\bar{\vx} (t) , t ; \bar{\vb} , \tfbar)  + V_{\vx}^{\T}  \delta \vx (t)   + V_{\vb}^{\T}  \delta \vb  +  V_{\tf}  \delta \tf   \bigg) \\
     - & \frac{ \rd}{ \rd t} \bigg(  \frac{1}{2} \deltaxbtfTo  \vMatrixVo \deltaxbtfo  \bigg)  \\
       + &   V_{\vx} ^{\T} F + \delta \vx (t) ^{\T} V_{\vx \vx} F + \delta \vb ^{\T} V_{\vb \vx} F + \delta \tf V_{\tf \vx} F \\
       + & \frac{1}{2} \deltaxbtfTo  \vMatrixVxxxFo \deltaxbtfo. 
       \label{eqn:HJB_lhs_expand}
                \end{split}   
       \end{align}
     The next step is to work with the expansion of the  right-hand side of the HJB equation in \eqref{eqn:HJB}. In particular, we have that
       \begin{align}
         V_{\vx} (\bar{\vx} (t) + \delta \vx  (t) , t ;   \bar{\vb} &+ \delta \vb , \tfbar + \delta \tf )     \approx ~   V_{\vx} (\bar{\vx}  (t) , t ; \bar{\vb} , \tfbar)  + V_{\vx \vx}  \delta \vx (t)   + V_{\vx \vb}  \delta \vb  +  V_{\vx \tf}  \delta \tf \nonumber \\
              &  + \frac{1}{2} \deltaxbtfTo  \vMatrixVxxxo \deltaxbtfo.
                \end{align}
           In addition, the running cost and the dynamics are expanded as follows:                
               \begin{align}    
       {\vL}(\vx(t),\vu(t), t)   =~ {\vL} ( \bar{\vx} (t)  &+ \delta {\vx} (t)  ,\bar{\vu} (t)  + \delta {\vu} (t)  , t ) \approx ~ \vL ( \bar{\vx} (t)  ,\bar{\vu} (t) , t ) + \vLx ^{\T} \delta \vx (t)  + \vLu ^{\T} \delta \vu (t) \nonumber  \\
          &+ \frac{1}{2}  \deltaxuTo \vMatrixLo \deltaxuo, 
                   \end{align}
               \begin{align}  
      F(\vx(t),\vu(t),t)  =~& F(\bar{\vx} (t)  + \delta \vx (t) ,\bar{\vu} (t)  + \delta \vu (t) ,t)  \approx ~  F(\bar{\vx} (t)  ,\bar{\vu} (t)  ,t)  + F_{\vx} \delta \vx (t)  +  F_{\vu} \delta \vu (t) .
       \end{align}
Therefore, the right hand side of (\ref{eqn:HJB}) can be expressed as 
\begin{align}
%& \min_{\delta \vu (t) } \bigg[ {\vL} \big( \bar{\vx} (t)  + \delta {\vx}  (t) ,\bar{\vu} (t)  + \delta {\vu} (t), t \big) +  V_{\vx} (\bar{\vx} (t) + \delta \vx (t)  , \bar{\vb} + \delta \vb , \tfbar + \delta \tf ;  t  ) F(\bar{\vx} (t)  + \delta \vx (t) ,\bar{\vu} (t)  + \delta \vu (t) ,t) \bigg] \nonumber \\
& \min_{\delta \vu (t) } \bigg\{ \vL ( \bar{\vx} (t)  ,\bar{\vu} (t), t   ) + \vLx ^{\T} \delta \vx (t)  + \vLu ^{\T} \delta \vu (t)   + \frac{1}{2}  \deltaxuTo \vMatrixLo \deltaxuo \nonumber \\
& +~{\color{black} V_{\vx}^{\T}  F } + V_{\vx}^{\T} F_{\vx} \delta \vx (t)  + V_{\vx}^{\T} F_{\vu} \delta \vu (t)  + ~{\color{black} \delta \vx (t)  ^{\T} V_{\vx \vx} F } +  \delta \vx (t)  ^{\T} V_{\vx \vx} F_{\vx} \delta \vx (t)  + \delta \vx (t)  ^{\T} V_{\vx \vx} F_{\vu} \delta \vu (t)  \nonumber \\ 
& + ~{\color{black} \delta \vb ^{\T} V_{\vb \vx} F } +  \delta \vb ^{\T} V_{\vb \vx} F_{\vx} \delta \vx (t)  + \delta \vb ^{\T} V_{\vb \vx} F_{\vu} \delta \vu (t)  + ~{\color{black} \delta \tf  V_{\tf \vx} F } +  \delta \tf  V_{\tf \vx} F_{\vx} \delta \vx (t)  + \delta \tf  V_{\tf \vx} F_{\vu} \delta \vu (t)  \nonumber \\
& + ~{\color{black} \frac{1}{2} \deltaxbtfTo  \vMatrixVxxxFo \deltaxbtfo } + H.O.T. \bigg\}.
    \label{eqn:HJB_rhs_expand}
\end{align}       
       Equating (\ref{eqn:HJB_lhs_expand}) with (\ref{eqn:HJB_rhs_expand}) and cancel like terms, we get 
       \begin{align}
& -\frac{ \rd}{ \rd t} \bigg(  V  + V_{\vx}^{\T}  \delta \vx (t)  + V_{\vb}^{\T}  \delta \vb  +  V_{\tf}  \delta \tf  + \frac{1}{2} \deltaxbtfTo  \vMatrixVo \deltaxbtfo  \bigg) =  \nonumber \\
& \min_{\delta \vu (t) } \bigg\{ \vL + \vLx ^{\T} \delta \vx (t)  + \vLu ^{\T} \delta \vu (t)   + \frac{1}{2}  \deltaxuTo \vMatrixLo \deltaxuo + V_{\vx}^{\T} F_{\vx} \delta \vx (t)  + V_{\vx}^{\T} F_{\vu} \delta \vu (t)  \nonumber \\
&  +  \delta \vx  (t) ^{\T} V_{\vx \vx} F_{\vx} \delta \vx (t)  + \delta \vx (t)  ^{\T} V_{\vx \vx} F_{\vu} \delta \vu (t)   +  \delta \vb ^{\T} V_{\vb \vx} F_{\vx} \delta \vx (t)  + \delta \vb ^{\T} V_{\vb \vx} F_{\vu} \delta \vu (t) \nonumber \\
& +  \delta \tf  V_{\tf \vx} F_{\vx} \delta \vx (t)  + \delta \tf  V_{\tf \vx} F_{\vu} \delta \vu (t)  \bigg\}.
		\label{eqn:HJB_expand}
       \end{align}
       
       To find the $\delta \vu(t)$ that minimize the equation, we take derivative of the right hand side of (\ref{eqn:HJB_expand}) and set it to $0$,  
       \begin{align}
       0 = \vL_{\vu} + \vL_{\vu \vu} \delta \vu (t) + (\frac{1}{2} \vL_{\vu \vx} + \frac{1}{2} \vL_{\vx \vu}^{\T} + F_{\vu}^{\T} V_{\vx \vx}) \delta \vx (t) + F_{\vu}^{\T} V_{\vx} + F_{\vu}^{\T} V_{\vx \vb} \delta \vb + F_{\vu}^{\T} V_{\vx \tf} \delta \tf.
       \end{align}
       The update law for the control is thus given by 
       \begin{align}
       \delta \vu (t)  = {\vl (t)} + {\vK_{\vx} (t)} \delta \vx (t)    + {\vK_{\vb} (t)} \delta \vb + {\vK_{\tf} (t)} \delta \tf.
       \label{eqn:opt_du_cont}
       \end{align}
       where the terms $  \vl (t),  \vK_{\vx}(t),\vK_{\vb} (t)   $  and $ \vK_{\tf} (t) $ are defined as follows
       \begin{align}
       \vl (t) = - \vL_{\vu \vu}^{-1} (\vL_{\vu} + F_{\vu}^{\T} V_{\vx}), & \quad \vK_{\vx} (t) = - \vL_{\vu \vu}^{-1} (\frac{1}{2} \vL_{\vu \vx} + \frac{1}{2} \vL_{\vx \vu}^{\T} + F_{\vu}^{\T} V_{\vx \vx} ), \nonumber \\
        \vK_{\vb} (t) = - \vL_{\vu \vu}^{-1} F_{\vu}^{\T} V_{\vx \vb} , & \quad \vK_{\tf} (t) = - \vL_{\vu \vu}^{-1} F_{\vu}^{\T} V_{\vx \tf}.
        \label{eqn:du_cont_coeffs}
       \end{align}
       Note that $\vL_{\vu \vu}$ is guaranteed to be invertible if the running cost $\vL = g(\vx) + \vu^{\T} R \vu$, where $R > 0$. This type of cost is normal for a mechanical system where we would like to minimize the energy cost of the control.

      Substitution of the  optimal policy variation $ \delta \vu $ back to the HJB  equation results in a set of backward ordinary differential equations that propagate the expansion of the value function  which  consists of the terms  $V, V_{\vb},V_{\vx \vx},V_{\vb \vb},V_{\vx \vb},V_{\vx \tf} $   and  $V_{\vb \tf} $. These backward differential equations  are given as follows     
        \begin{align}\label{eqn:Backward_pass}\begin{split}
      -\frac{ \rd}{ \rd t} V &= \vL - \frac{1}{2} \vl^{\T} \vL_{\vu \vu} \vl ,\\  -\frac{ \rd}{ \rd t} V_{\vx} &= \vLx - \vK_{\vx}^{\T} \vL_{\vu \vu} \vl + F_{\vx}^{\T} V_{\vx} , \\
      -\frac{ \rd}{ \rd t} V_{\vb} &= \vK_{\vb}^{\T} \vL_{\vu} , \\    -\frac{ \rd}{ \rd t} V_{\tf} &= \vK_{\tf}^{\T} \vL_{\vu} , \\
      -\frac{ \rd}{ \rd t} V_{\vx \vx} &= \vL_{\vx \vx} - \vK_{\vx}^{\T} \vL_{\vu \vu} \vK_{\vx}  + 2 F_{\vx}^{\T} V_{\vx \vx} , \\
      -\frac{ \rd}{ \rd t} V_{\vb \vb} &= - \vK_{\vb}^{\T} \vL_{\vu \vu} \vK_{\vb} , \\     -\frac{ \rd}{ \rd t} V_{\tf \tf} &= - \vK_{\tf}^{\T} \vL_{\vu \vu} \vK_{\tf} , \\
       -\frac{ \rd}{ \rd t} V_{\vx \vb} &= \vL_{\vx \vu} \vK_{\vb}  + F_{\vx}^{\T} V_{\vx \vb} + V_{\vx \vx} F_{\vu} \vK_{\vb} , \\
       -\frac{ \rd}{ \rd t} V_{\vx \tf} &= \vL_{\vx \vu} \vK_{\tf}  + F_{\vx}^{\T} V_{\vx \tf}  + V_{\vx \vx} F_{\vu} \vK_{\tf} , \\
      -\frac{ \rd}{ \rd t} V_{\vb \tf} &=  \vK_{\vb}^{\T} F_{\vu}^{\T} V_{\vx \tf} ,
      \end{split}
       \end{align}
       where all the quantities are evaluated at $(\bar{\vx} (t), \bar{\vu} (t), \bar{\vb}, \tfbar)$.  To numerically solve  the  equations in \eqref{eqn:Backward_pass} one has to compute the terminal conditions.  In the next section we  present the derivation for  the terminal condition and provide an overview of the algorithm. 
       
       \subsection{Terminal Conditions}
       
       The terminal conditions can be determined by the following procedure. 
       
              From  
                \begin{equation}
                   V(\vx(t_{0}), t_{0}; \vb, \tf)  = \min_{\vu (\cdot)}J(\vx (\cdot),\vu (\cdot))  =  \min_{\vu (\cdot)} \bigg\{ \int_{t_{0}}^{\tf} {\vL}(\vx(t),\vu(t))  \rd t  +  \Phi(x(\tf), \vb, \tf) \bigg\},
                  \end{equation}
                  we have that for any $t \in [t_0, t_f]$, 
                    \begin{equation}
                       V(\vx(t), t; \vb, \tf)  = \min_{\vu (\cdot)}J(\vx (\cdot),\vu (\cdot))  =  \min_{\vu (\cdot)} \bigg\{ \int_{t}^{\tf} {\vL}(\vx(s),\vu(s))  \rd s + \Phi(x(\tf), \vb, \tf) \bigg\}.
                      \end{equation}
                      Therefore, 
       \begin{align}
      & V(\bar{\vx} ( \tfbar) + \delta \vx ( \tfbar), \tfbar; \bar{\vb} + \delta \vb, \tfbar + \delta \tf) \nonumber \\
       = & \min_{\vu (\cdot)} \bigg\{ \int_{\tfbar}^{\tfbar + \delta \tf} \vL(\vx (t), \vu (t), t )  \rd t + \Phi(\vx(\tfbar + \delta \tf), \bar{\vb} + \delta \vb, \tfbar + \delta \tf)  \bigg\} \nonumber \\
            \end{align}
          \begin{align}
        &  \vL (\bar{\vx} (\tfbar), \bar{\vu}(\tfbar), t )  \delta \tf + \Phi(\bar{\vx}(\tfbar) + \delta \vx(\tfbar) + \dot{\bar{\vx}} (\tfbar) \delta \tf, \bar{\vb} + \delta \vb, \tfbar + \delta \tf)  \nonumber \\
       &\approx  ~\vL (\bar{\vx} (\tfbar), \bar{\vu}(\tfbar), t )  \delta \tf + \Phi(\bar{\vx} (\tfbar), \bar{\vb}, \tfbar) + \Phi_{\vx} ^{\T} (\delta \vx (\tfbar) + F (\bar{\vx} (\tfbar), \bar{\vu} (\tfbar), \tfbar) \delta \tf) \nonumber \\
       & + \Phi_{\vb} (\bar{\vx} (\tfbar), \bar{\vu} (\tfbar), \tfbar) ^{\T} \delta \vb  + \Phi_{\tf} (\bar{\vx} (\tfbar), \bar{\vu} (\tfbar), \tfbar) \delta \tf   \nonumber \\ 
       &+ \frac{1}{2} \deltaxbtfend^{\T} \vMatrixPhiend \deltaxbtfend 
             \end{align}
          \begin{align}
           &V(\bar{\vx} ( \tfbar) + \delta \vx ( \tfbar), \tfbar; \bar{\vb} + \delta \vb, \tfbar + \delta \tf)  =  ~ \Phi + \Phi_{\vx} ^{\T} \delta \vx + \Phi_{\vb} ^{\T} \delta \vb  + ( \vL + \Phi_{\vx} ^{\T} F + \Phi_{\tf} ) \delta \tf \nonumber \\
              & + \frac{1}{2} \deltaxbtfTo \begin{bmatrix} \Phi_{\vx \vx}  &  \Phi_{\vx \vb}  &   \Phi_{\vx \tf} + \Phi_{\vx \vx} F  \\ \Phi_{\vb \vx} &  \Phi_{\vb \vb}  &   \Phi_{\vb \tf} + \Phi_{\vb \vx} F \\ \Phi_{\tf \vx} + F^{\T} \Phi_{\vx \vx}  &  \Phi_{\tf \vb} + F^{\T} \Phi_{\vx \vb}   &   \Phi_{\tf \tf}  + 2 \Phi_{\tf \vx} F  + F^{\T} \Phi_{\vx \vx} F \end{bmatrix} \deltaxbtfo ,
       \end{align}
       where $\vx(\tfbar + \delta \tf)$ is evaluated by $\bar{\vx} (\tfbar) + \delta \vx (\tfbar)  + \dot{\bar{\vx}} (\tfbar)  \delta \tf$ and $\dot{\bar{\vx}} (\tfbar) = F(\bar{\vx} (\tfbar), \bar{\vu} (\tfbar), \tfbar).$ The arguments of the functions in the last line of equations are the same as those in the previous equations and are thus omitted. The minimization with respect to $\vu$ is dropped on the third line of equations because $\int_{\tfbar}^{\tfbar + \delta \tf} \vL(\vx (t), \vu (t))  \rd t$ is evaluated by $\vL (\bar{\vx} (\tfbar), \bar{\vu}(\tfbar))  \delta \tf$ and the latter is only a function of the nominal control.  Note that this approximation is relatively rough, since we approximate $\int_{\tfbar}^{\tfbar + \delta \tf} \vL(\vx (t), \vu (t))  \rd t$ by $\vL (\bar{\vx} (\tfbar), \bar{\vu}(\tfbar))  \delta \tf$ instead of  $\vL ({\vx} (\tfbar), {\vu}(\tfbar))  \delta \tf$ and follow up with an expansion on ${\vx} (\tfbar)$ and ${\vu}(\tfbar)$. But the simulation results suggest that such level of approximation is good enough. 
            
       Hence, at $t = \tfbar$, the terminal conditions are
              \begin{align}\begin{split}
              V(\bar{\vx} (\tfbar), \tfbar ; \bar{\vb}, \tfbar) &= \Phi(\bar{\vx} (\tfbar), \bar{\vb}, \tfbar) , \\
              V_{\vx}(\bar{\vx} (\tfbar), \tfbar ; \bar{\vb}, \tfbar) &= \Phi_{\vx}(\bar{\vx} (\tfbar), \bar{\vb}, \tfbar) , \\
              V_{\vb}(\bar{\vx} (\tfbar), \tfbar ; \bar{\vb}, \tfbar) &= \Phi_{\vb}(\bar{\vx} (\tfbar), \bar{\vb}, \tfbar) , \\
               V_{\tf}(\bar{\vx} (\tfbar), \tfbar ; \bar{\vb}, \tfbar) &= \vL(\bar{\vx} (\tfbar), \bar{\vu}(\tfbar)) + \Phi_{\vx}(\bar{\vx} (\tfbar), \bar{\vb}, \tfbar) ^{\T} F(\bar{\vx} (\tfbar), \bar{\vu}(\tfbar)) + \Phi_{\tf}(\bar{\vx} (\tfbar), \bar{\vb}, \tfbar) , \\
              V_{\vx \vx}(\bar{\vx} (\tfbar), \tfbar ; \bar{\vb}, \tfbar) &= \Phi_{\vx \vx}(\bar{\vx} (\tfbar), \bar{\vb}, \tfbar) , \\
              V_{\vb \vb}(\bar{\vx} (\tfbar), \tfbar ; \bar{\vb}, \tfbar) &= \Phi_{\vb \vb}(\bar{\vx} (\tfbar), \bar{\vb}, \tfbar) , \\
              V_{\tf \tf}(\bar{\vx} (\tfbar), \tfbar ; \bar{\vb}, \tfbar) &= \Phi_{\tf \tf}(\bar{\vx} (\tfbar), \bar{\vb}, \tfbar)  + 2 \Phi_{\tf \vx}(\bar{\vx} (\tfbar), \bar{\vb}, \tfbar) F(\bar{\vx} (\tfbar), \bar{\vu}(\tfbar))  \\
              & + F(\bar{\vx} (\tfbar), \bar{\vu}(\tfbar))^{\T} \Phi_{\vx \vx}(\bar{\vx} (\tfbar), \bar{\vb}, \tfbar) F(\bar{\vx} (\tfbar), \bar{\vu}(\tfbar)) ,\\                                      
              V_{\vx \vb}(\bar{\vx} (\tfbar), \tfbar ; \bar{\vb}, \tfbar) &= \Phi_{\vx \vb}(\bar{\vx} (\tfbar), \bar{\vb}, \tfbar) , \\
              V_{\vx \tf}(\bar{\vx} (\tfbar), \tfbar ; \bar{\vb}, \tfbar) &= \Phi_{\vx \tf}(\bar{\vx} (\tfbar), \bar{\vb}, \tfbar) + \Phi_{\vx \vx}(\bar{\vx} (\tfbar), \bar{\vb}, \tfbar) F(\bar{\vx} (\tfbar), \bar{\vu}(\tfbar)) , \\
              V_{\vb \tf}(\bar{\vx} (\tfbar), \tfbar ; \bar{\vb}, \tfbar) &=  \Phi_{\vb \tf}(\bar{\vx} (\tfbar), \bar{\vb}, \tfbar) + \Phi_{\vb \vx}(\bar{\vx} (\tfbar), \bar{\vb}, \tfbar) F (\bar{\vx} (\tfbar), \bar{\vu}(\tfbar)) .     
                \end{split}   
              \end{align}

       Given the boundary conditions of the value function and its derivatives, we can back-propagate the differential equations we derived earlier to find their values. Our next step is to update the control through (\ref{eqn:opt_du_cont}), and in order to do so, we need to find the update law of $\delta \vb$ and $\delta \tf$.

        We follow the derivation in \cite{jacobson1970differential} and set 
                                   \begin{align} \label{eqn:update_btf2}
                                   \begin{bmatrix} \delta \vb \\ \delta \tf \end{bmatrix} = - \zeta \begin{bmatrix}  V_{\vb \vb} (\bar{\vx} (t_0), t_0; \bar{\vb}, \tfbar )  &   V_{\vb \tf} (\bar{\vx} (t_0), t_0; \bar{\vb}, \tfbar ) \\  V_{\tf \vb} (\bar{\vx} (t_0), t_0; \bar{\vb}, \tfbar )  &   V_{\tf \tf} (\bar{\vx} (t_0), t_0; \bar{\vb}, \tfbar ) \end{bmatrix} ^{-1}  \begin{bmatrix} V_{\vb} (\bar{\vx} (\tfbar), \tfbar ; \bar{\vb}, \tfbar) \\ V_{\tf} (\bar{\vx} (\tfbar), \tfbar ; \bar{\vb}, \tfbar) \end{bmatrix},
                                   \end{align}
                                   where $\zeta \in [0, 1]$ is introduced to ensure that the update of $\delta \vb$ and $\delta \tf$ are not too large.

       \section{Simulation Results}
       
           \subsection{Double Integrator}
\begin{figure}
                      \centering
                      \begin{subfigure}[b]{0.22\textwidth}
                              \includegraphics[width=\textwidth]{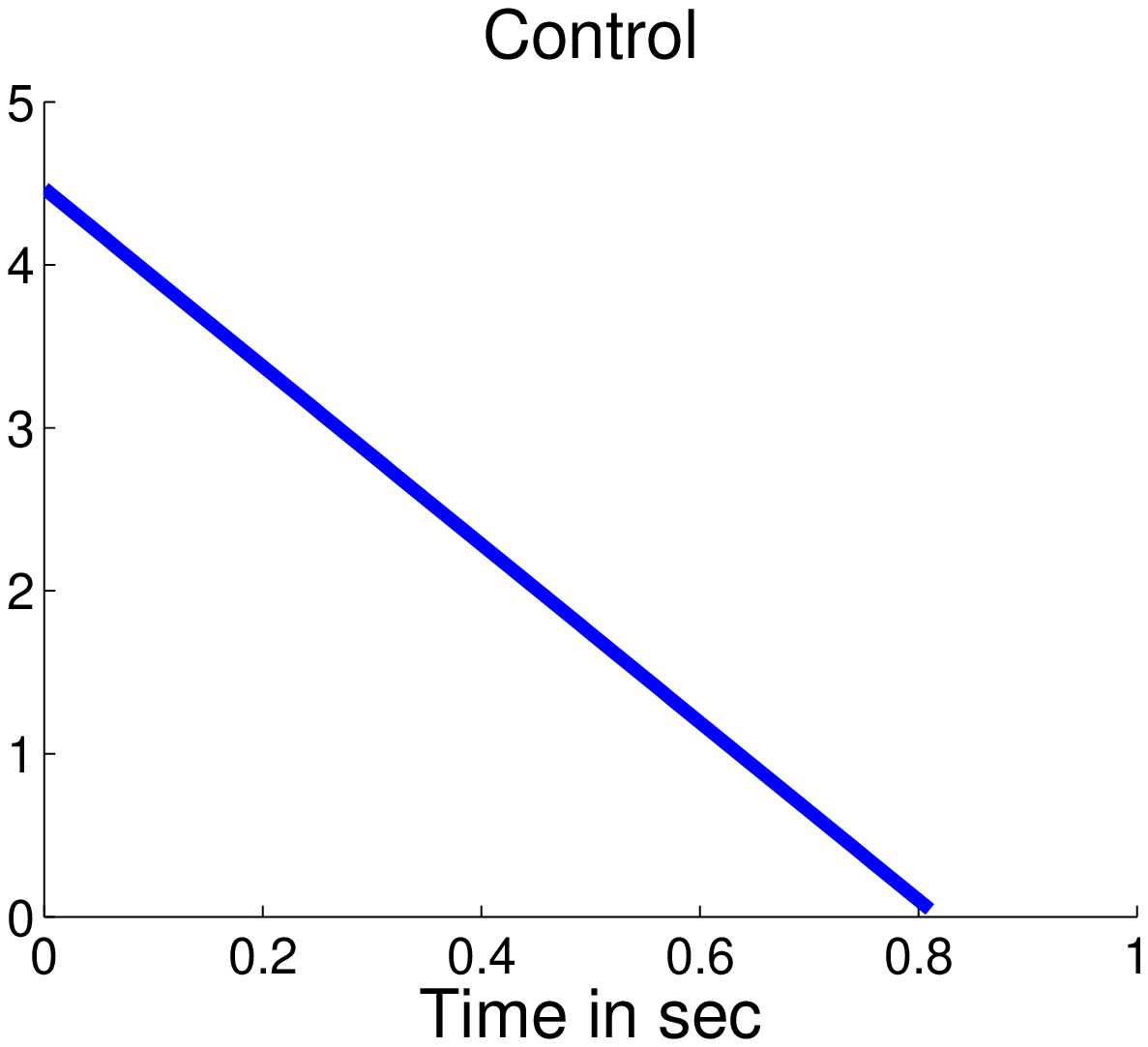}
                              \caption{R=0.1}
                              \label{fig:oc01}
                      \end{subfigure}%
                      ~ %add desired spacing between images, e. g. ~, \quad, \qquad, \hfill etc.
                        %(or a blank line to force the subfigure onto a new line)
                        \quad 
                      \begin{subfigure}[b]{0.22\textwidth}
                              \includegraphics[width=\textwidth]{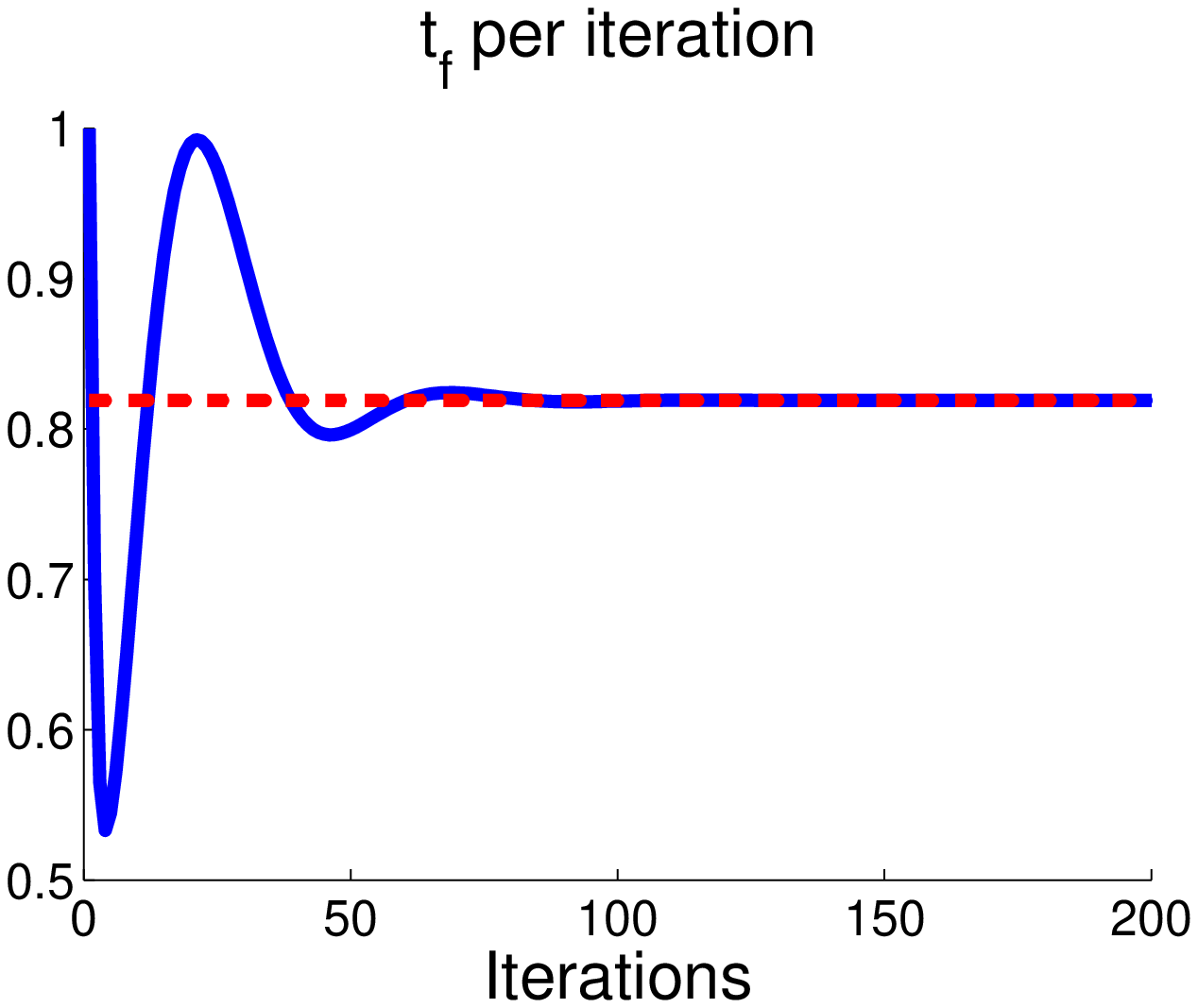}
                              \caption{R=0.1}
                              \label{fig:tf01}
                      \end{subfigure} 
                      ~ %add desired spacing between images, e. g. ~, \quad, \qquad, \hfill etc.
                        %(or a blank line to force the subfigure onto a new line)
                        \quad
                   \begin{subfigure}[b]{0.22\textwidth}
                              \includegraphics[width=\textwidth]{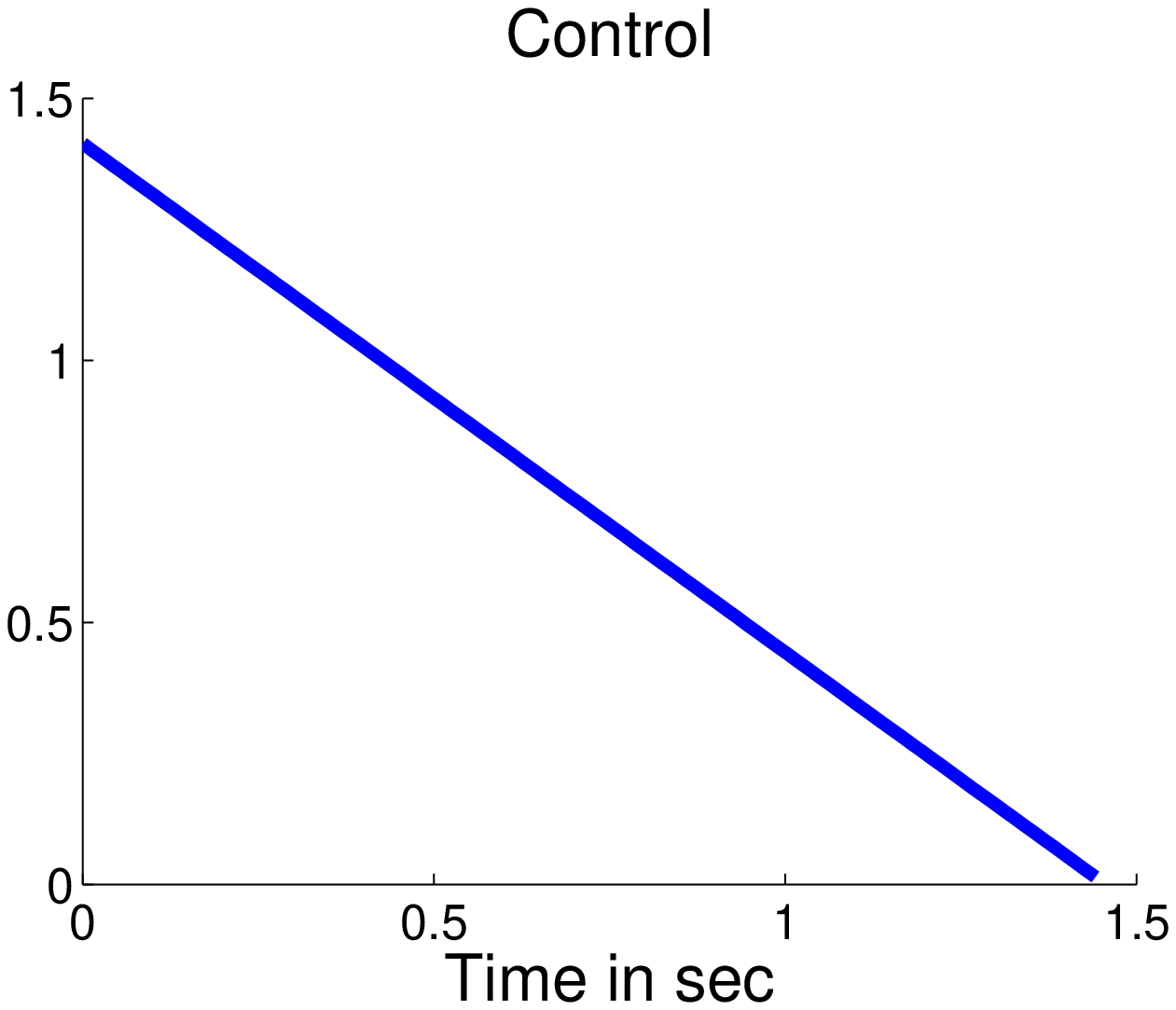}
                              \caption{R=1}
                              \label{fig:oc1}
                      \end{subfigure}\\
                        \begin{subfigure}[b]{0.22\textwidth}
                              \includegraphics[width=\textwidth]{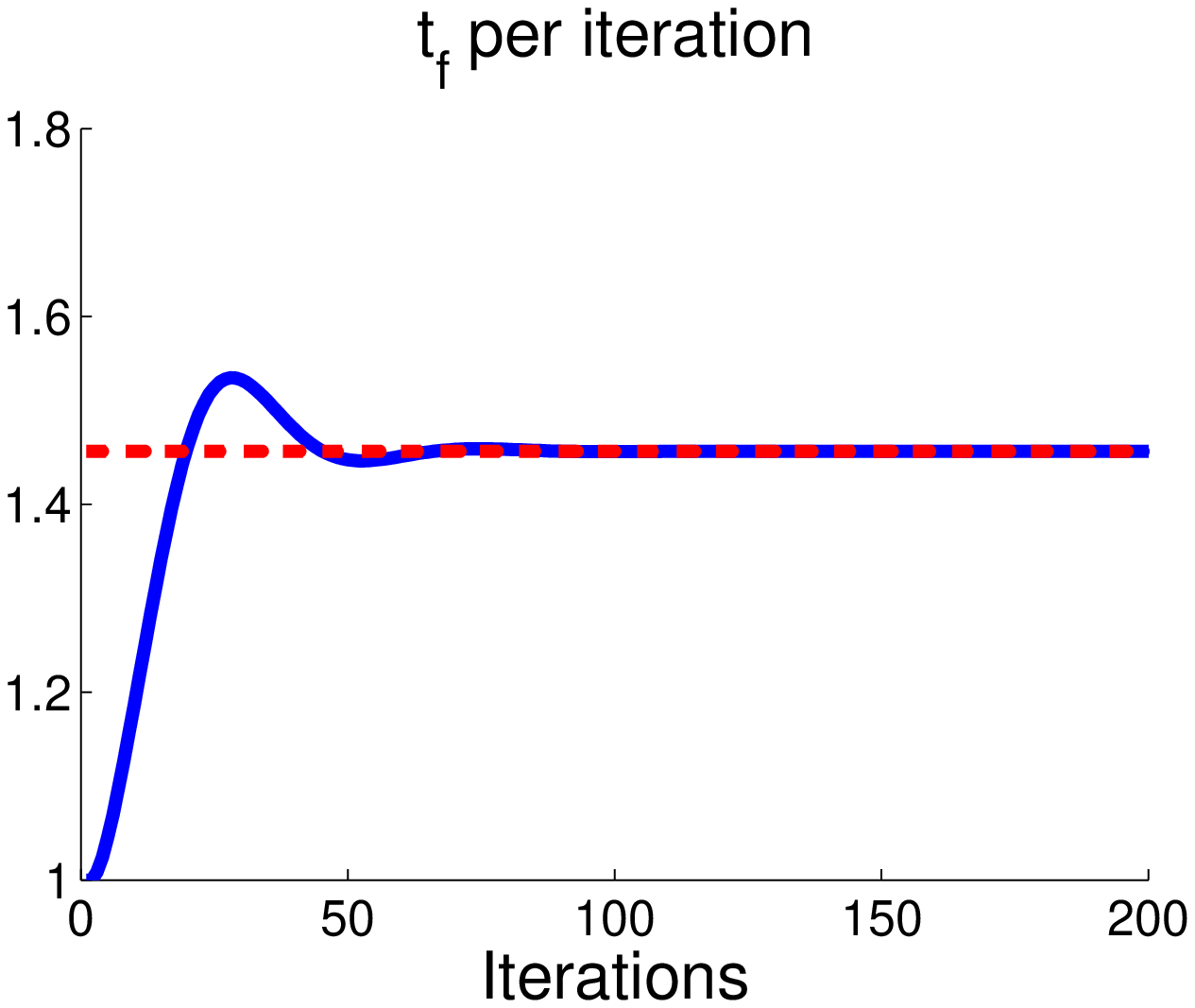}
                              \caption{R=1}
                              \label{fig:tf1}
                      \end{subfigure} 
                      \quad
                              \begin{subfigure}[b]{0.22\textwidth}
                              \includegraphics[width=\textwidth]{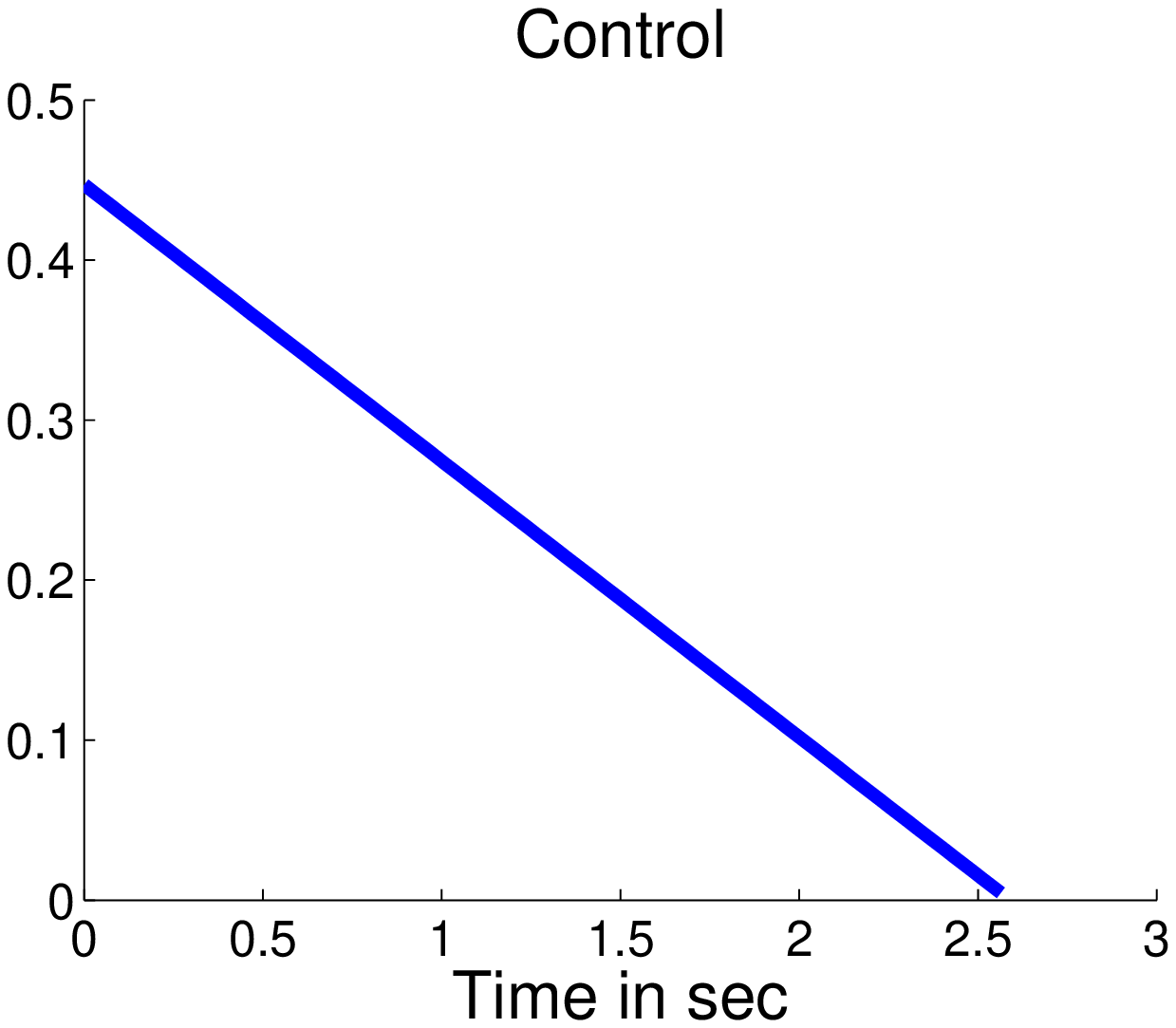}
                              \caption{R=10}
                              \label{fig:oc10}
                      \end{subfigure}
                      \quad
                              \begin{subfigure}[b]{0.22\textwidth}
                              \includegraphics[width=\textwidth]{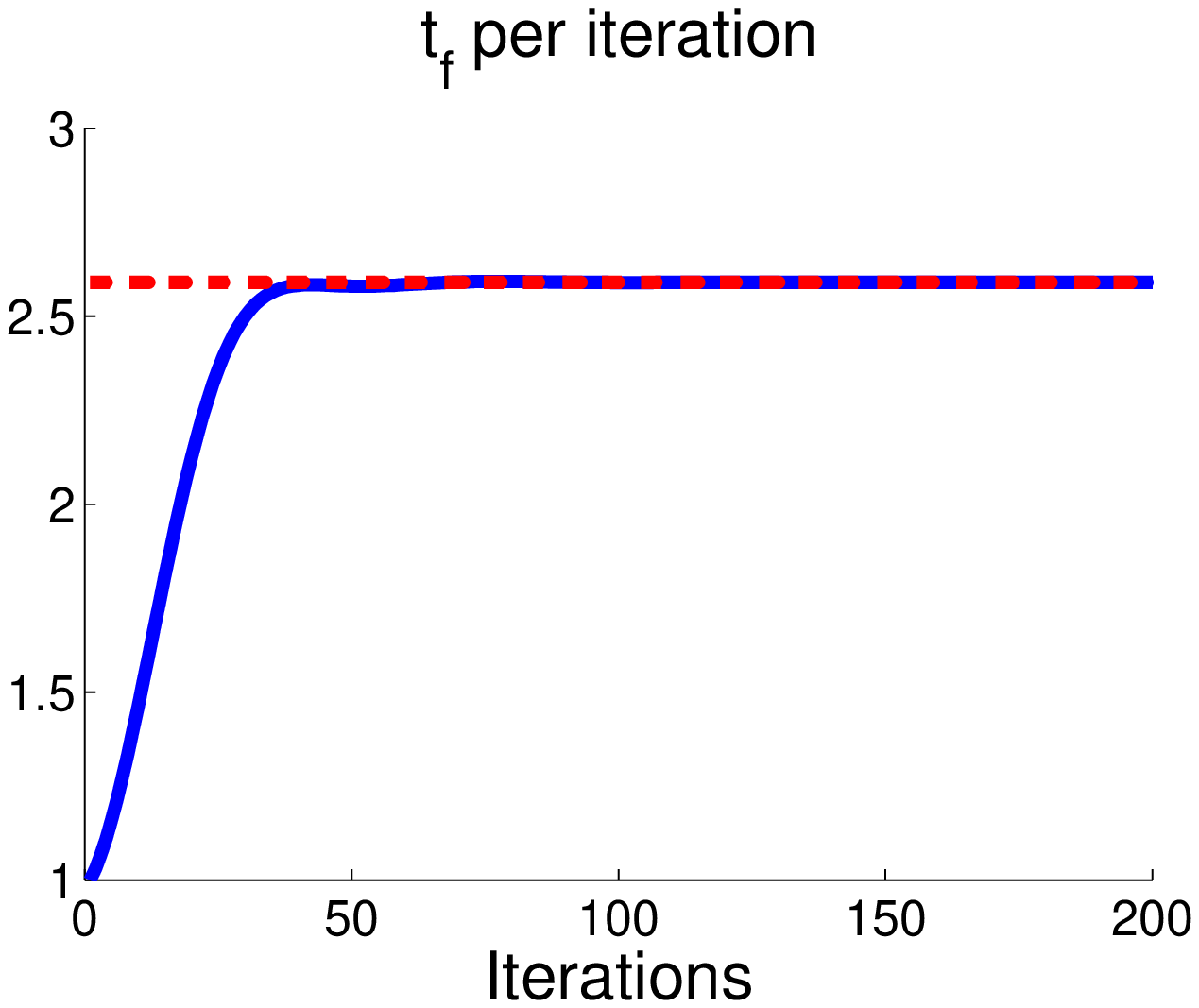}
                              \caption{R=10}
                              \label{fig:tf10}
                      \end{subfigure}
                      \caption{Figure \ref{fig:oc01}, \ref{fig:oc1}, \ref{fig:oc10} show the numerical optimal control for the cases when $R = 0.1, 1, 10$, respectively. Figure \ref{fig:tf01}, \ref{fig:tf1}, \ref{fig:tf10} present the $t_f$ per iteration for the cases when $R = 0.1, 1, 10$, respectively. Dashed red lines represents the according analytical optimal $t_f$.}
                      \label{fig:double_integrator}
              \end{figure}

       We first apply the algorithm on a simple system, namely, the double integrator. We compare our numerical result with the analytical solution to verify the algorithm. The dynamics is given by $ 
       \dot{x}_1  = x_2, $ and $    \dot{x}_2  = \vu. $ 
       Initial condition is $\vx(0) = [x_1(0); x_2(0)] = [0; 0]$.
       The cost $J = \int_{0}^{\tf} (1 + R \vu^2) \rd t.$ Terminal constraint is $x_1 - 1 = 0$. Introducing the Lagrange multiplier $\vb$, the cost can be reformulated as $
       J = \vb (x_1 - 1) + \int_{0}^{\tf} (1 + \frac{1}{2} R \vu^2) \rd t. $
       Given different values of $R$, we can find different optimal cost and terminal time. In particular, let $R = 0.1, 1, 10$, the corresponding terminal times are $  0.819, 1.456, 2.590$, respectively. Optimal control and $\tf$ per iteration when $R = 0.1, 1, 10$ are shown in Figure \ref{fig:double_integrator}.     
       
       Now we solve this problem analytically to verify the simulation results. Denote the co-states by $\lambda = [\lambda_1; \lambda_2]$, the Hamiltonian is given by 
       \begin{align}
       H = 1 + \frac{1}{2} R \vu^2 + \lambda_1 x_2 + \lambda_2 u.
       \end{align}
       The co-states satisfy the adjoint equations
       \begin{align}
       \dot{\lambda}_1 = - \fracpartial{H}{x_1} = 0, \\
       \dot{\lambda}_2 = - \fracpartial{H}{x_2} = - \lambda_1.
       \end{align}
       Utilizing the Pontryagin's minimum principle, the optimal control $\vu^*$ can be calculated from $ 
       0 = \fracpartial{H}{u} = R \vu + \lambda_2. $
       Hence, $\vu^* = -\lambda_2/R$. 
       Transversality conditions are such that $\lambda_1(\tf) = \nu$, $\lambda_2(\tf) = 0$, $H(\tf) = 0$. Given the previous information, we are ready to solve the problem.
       From $\dot{\lambda}_1 = 0$ and $\lambda_1(\tf) = \nu$, we get $$\lambda_1(t) \equiv \nu, t \in [0, \tf].$$ 
       Then from $\dot{\lambda}_2 = - \lambda_1$ and $\lambda_2(\tf) = 0$, we have $ \lambda_2(t) = \nu (\tf - t).$ 
       Therefore, 
       \begin{align}
        \vu^* = - \lambda_2/R = \frac{\nu}{R} (t - \tf).
       \end{align}
       Note that the optimal control is a linear function of $t$.   Furthermore, boundary conditions yields $\tf^* = (\frac{9}{2} R)^{\frac{1}{4}}$ and $\nu^* = - \frac{2}{3} \tf^* = - \frac{2}{3} (\frac{9}{2} R)^{\frac{1}{4}}.$  When $R = 0.1, 1, 10$, $\tf^* = 0.8190, 1.4565, 2.5900$, respectively, which is consistent with the numerical simulation results presented in the control plots in Figure \ref{fig:double_integrator}. 

\subsection{Cart Pole}
\begin{figure}[htb]%{0.45\textwidth}
                              	\centering
                              	\includegraphics[width=0.4\textwidth]{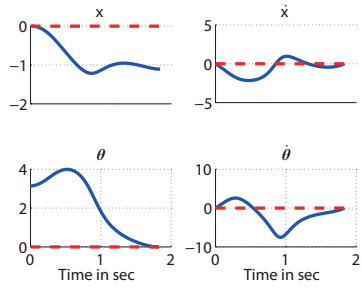}
                              	\caption{Optimal trajectories of the states in blue. Red lines represent the desired terminal states.}
                              	\label{fig:optimal_cart_traj}
                              \end{figure}

\begin{figure}
        \centering
         \begin{subfigure}[b]{0.3\textwidth}
                        \includegraphics[width=\textwidth]{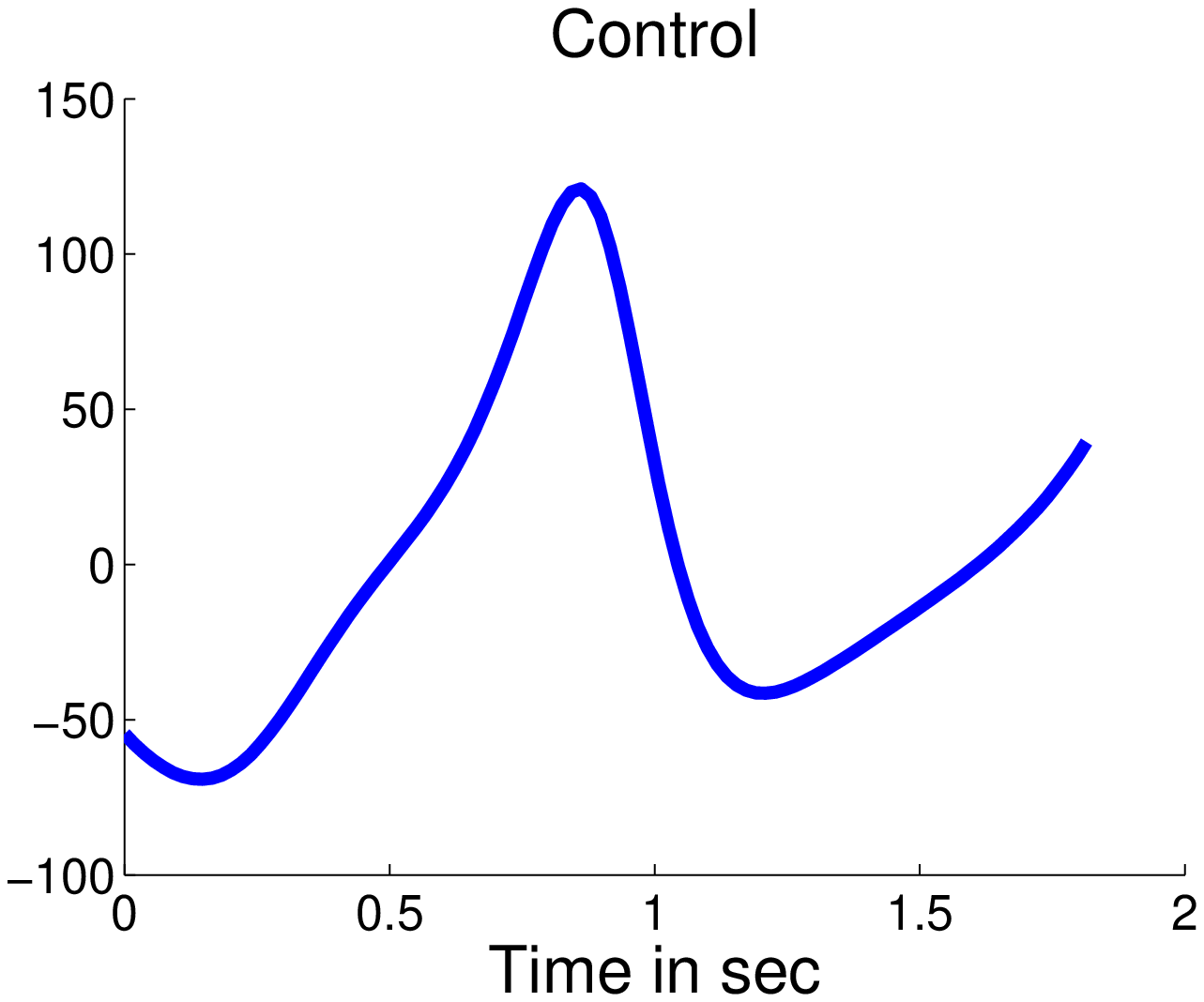}
                        \caption{Optimal control of the cart pole system.}
                        \label{fig:optimal_cart_u}
                \end{subfigure}%
                ~
        \begin{subfigure}[b]{0.3\textwidth}
                \includegraphics[width=\textwidth]{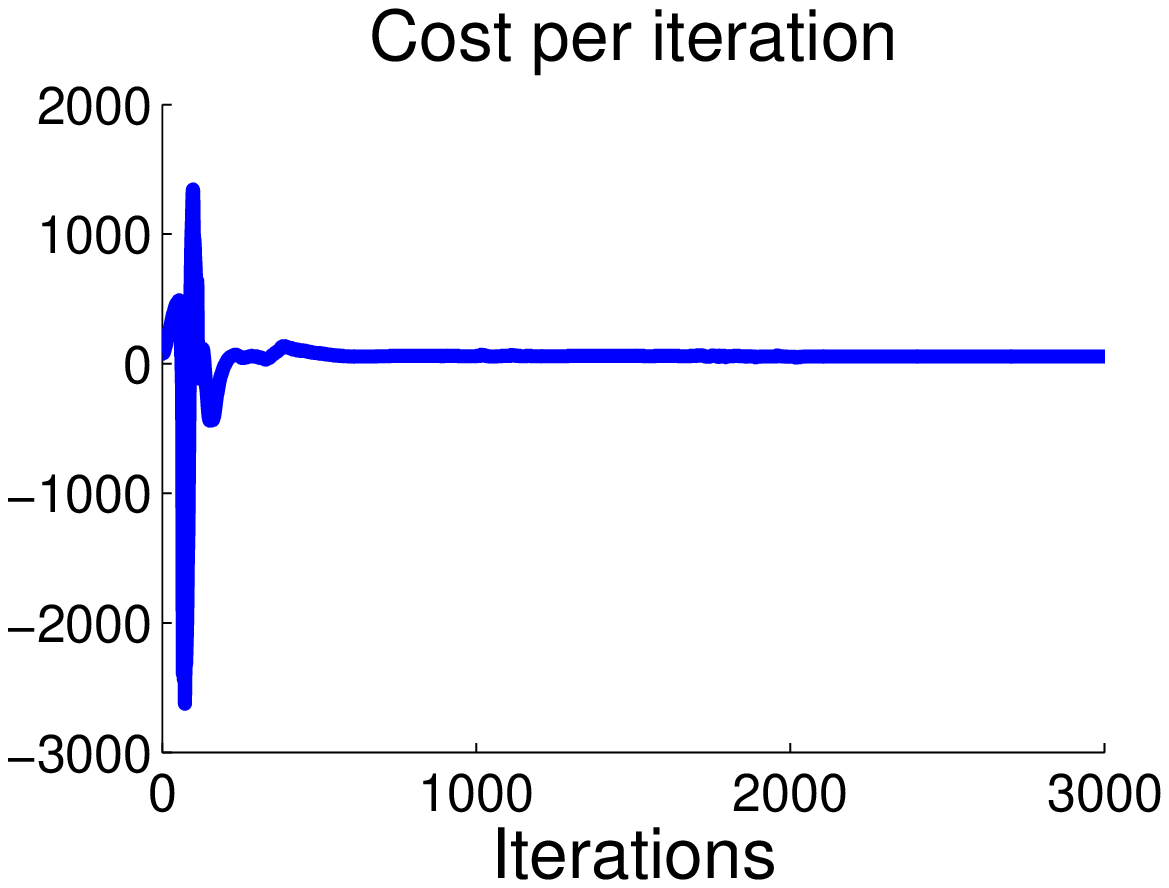}
                \caption{Cost per iteration.}
                \label{fig:cart_cost}
        \end{subfigure}%
        ~ %add desired spacing between images, e. g. ~, \quad, \qquad, \hfill etc.
          %(or a blank line to force the subfigure onto a new line)
        \begin{subfigure}[b]{0.3\textwidth}
                \includegraphics[width=\textwidth]{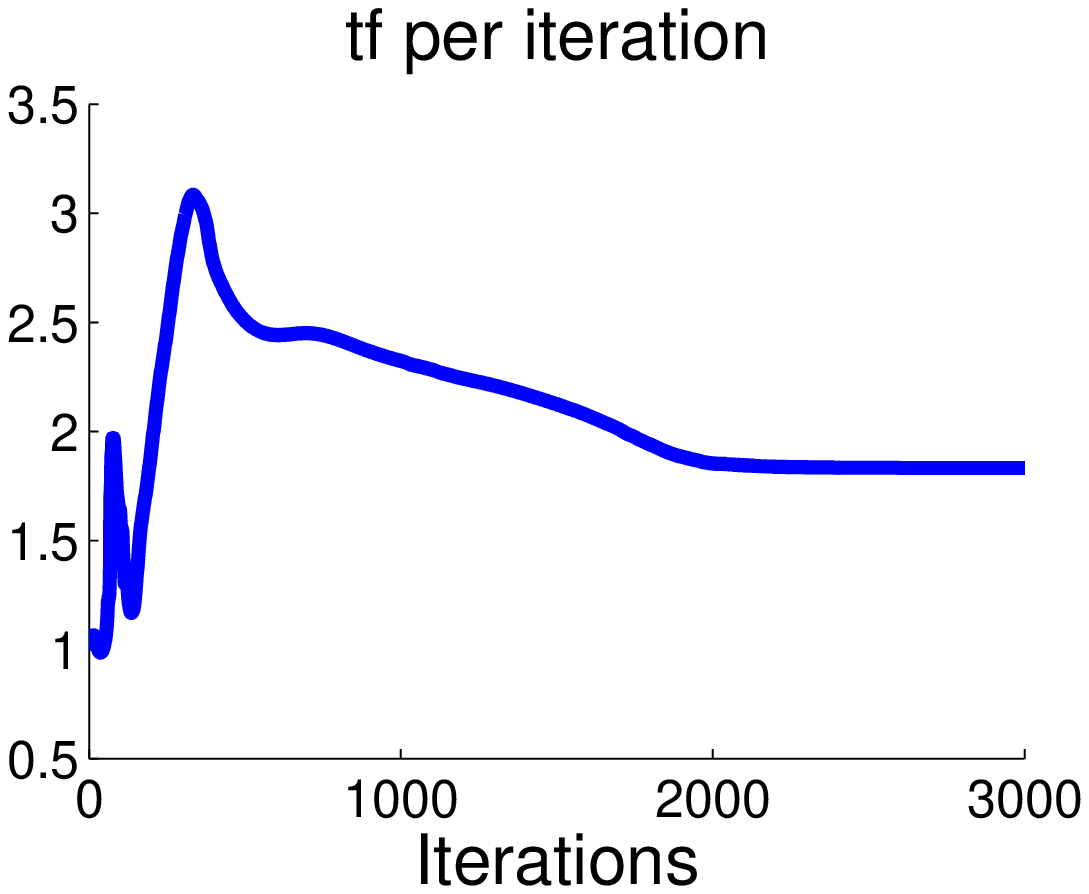}
                \caption{$t_f$ per iteration.}
                \label{fig:cart_tf}
        \end{subfigure}
        \caption{Cost and $t_f$ per iteration for the cart pole system.}\label{fig:cart_ctf}
\end{figure}

In this subsection, we apply our algorithm on the inverted pendulum on a cart, as known as the cart pole problem,
%\begin{align*}
%(M + m) \ddot{x} - ml \ddot{\theta} \cos \theta + ml^2 \dot{\theta}^2 \sin \theta = u, \\
%l\ddot{\theta} - g \sin \theta = \ddot{x} \cos \theta.
%\end{align*}
with  $M = 10$  the mass of the cart, $m = 1$  and $l = 0.5$ are the mass and length of the pendulum, $g = 9.8$ the gravitational acceleration and  $u$  the force applied to the cart. The state $\vx = [x, \dot{x}, \theta, \dot{\theta}]$. The goal is to bring the state from $\vx(0) = [0, 0, \pi, 0]$ to $p = [0, 0, 0 ,0]$, which represents the case where the pendulum is pointing strait up.  The cost function is given by 
\begin{align*}
J = \frac{1}{2} \int_{0}^{\tf} [ c_t + (\vx - p)^{\T} Q (\vx - p) + \vu^{\T} R_{\vu} \vu ] + \lambda^{\T} ([\vx_3 (t_f); \vx_4 (t_f)] - [p_3; p_4 ]),
\end{align*}
where $Q = \text{diag} \{ 0, 0, 1, 1\}$ and $R_{\vu} = 0.01$. Initial values are given as $\tf = 1$, $\lambda = [0, 0]^{\T}$. The multipliers $\gamma = 0.05$ and $\epsilon = 0.05$.  We run the algorithm for 300 iterations and the convergence is achieved at around 200th iteration. Figure  \ref{fig:optimal_cart_u} presents the optimal control $\vu^*$. The corresponding optimal trajectories of the states are depicted in Figure \ref{fig:optimal_cart_traj}. Cost and $t_f$ per iteration are shown in Figure \ref{fig:cart_cost} and \ref{fig:cart_tf}, respectively.

\subsection{Quadrotor}
\begin{figure}[htb]%{0.45\textwidth}
       	\centering
       	\includegraphics[width=0.8\textwidth]{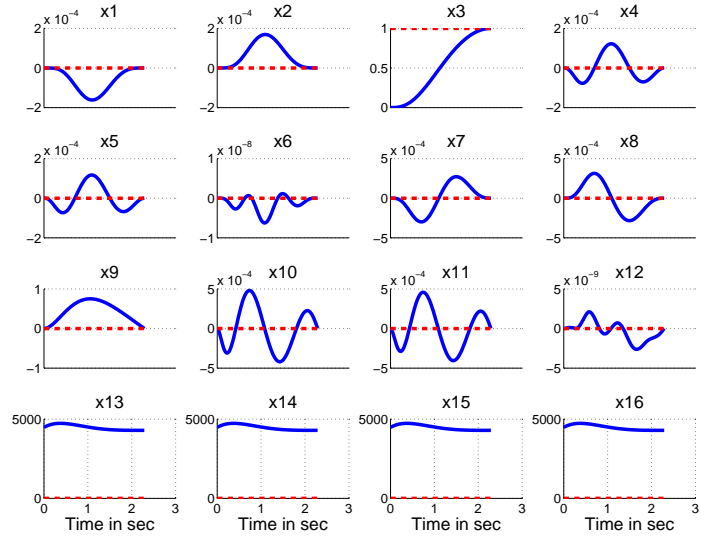}
       	\caption{Optimal trajectories of the states of the quadrotor in blue. Dashed red lines represent the desired terminal states.}
       	\label{fig:optimal_quad_traj}
       \end{figure}

\begin{figure}
        \centering
                \begin{subfigure}[b]{0.3\textwidth}
                        \includegraphics[width=\textwidth]{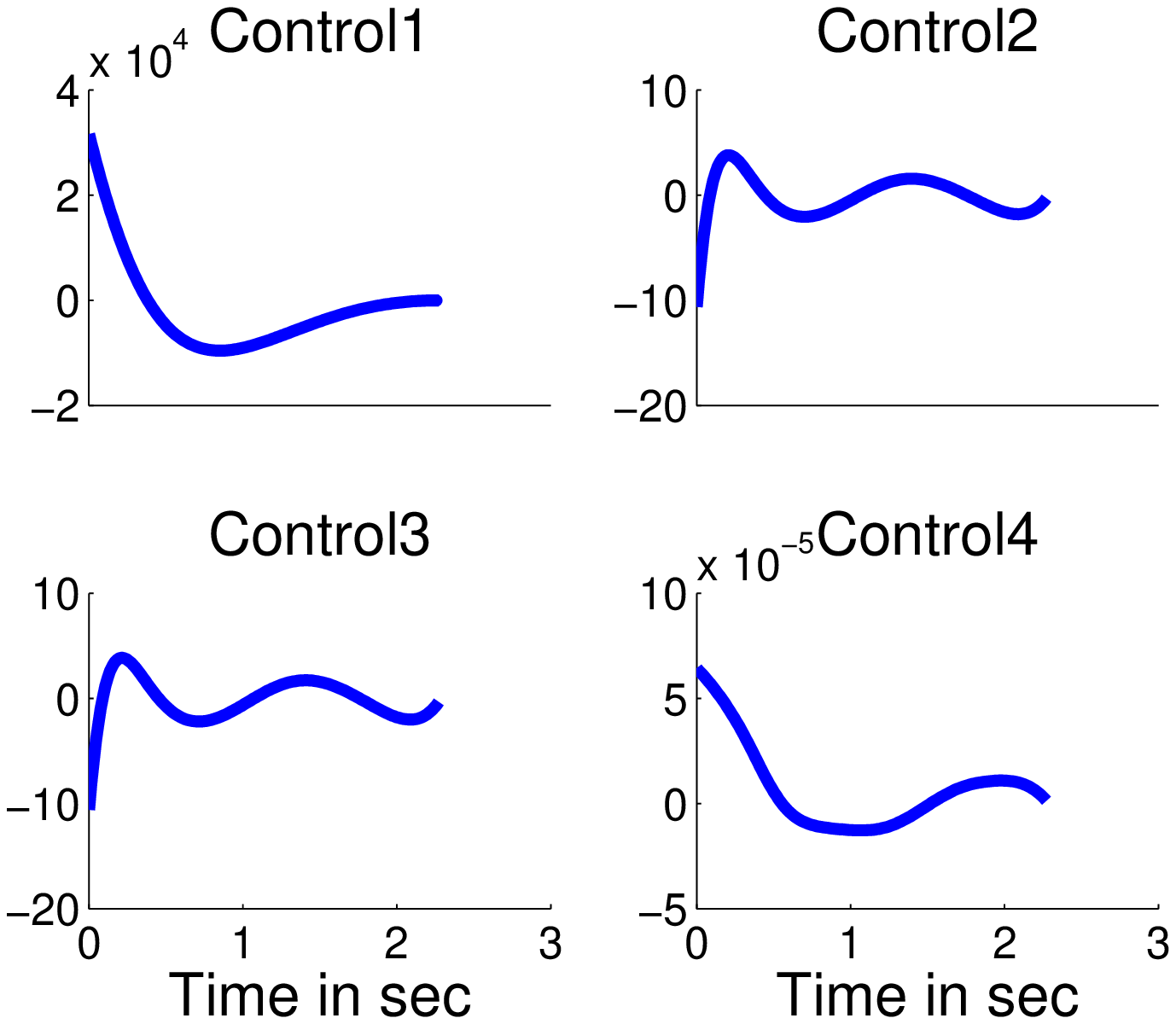}
                        \caption{Optimal controls.}
                        \label{fig:optimal_quad_u}
                \end{subfigure}%
                ~
        \begin{subfigure}[b]{0.3\textwidth}
                \includegraphics[width=\textwidth]{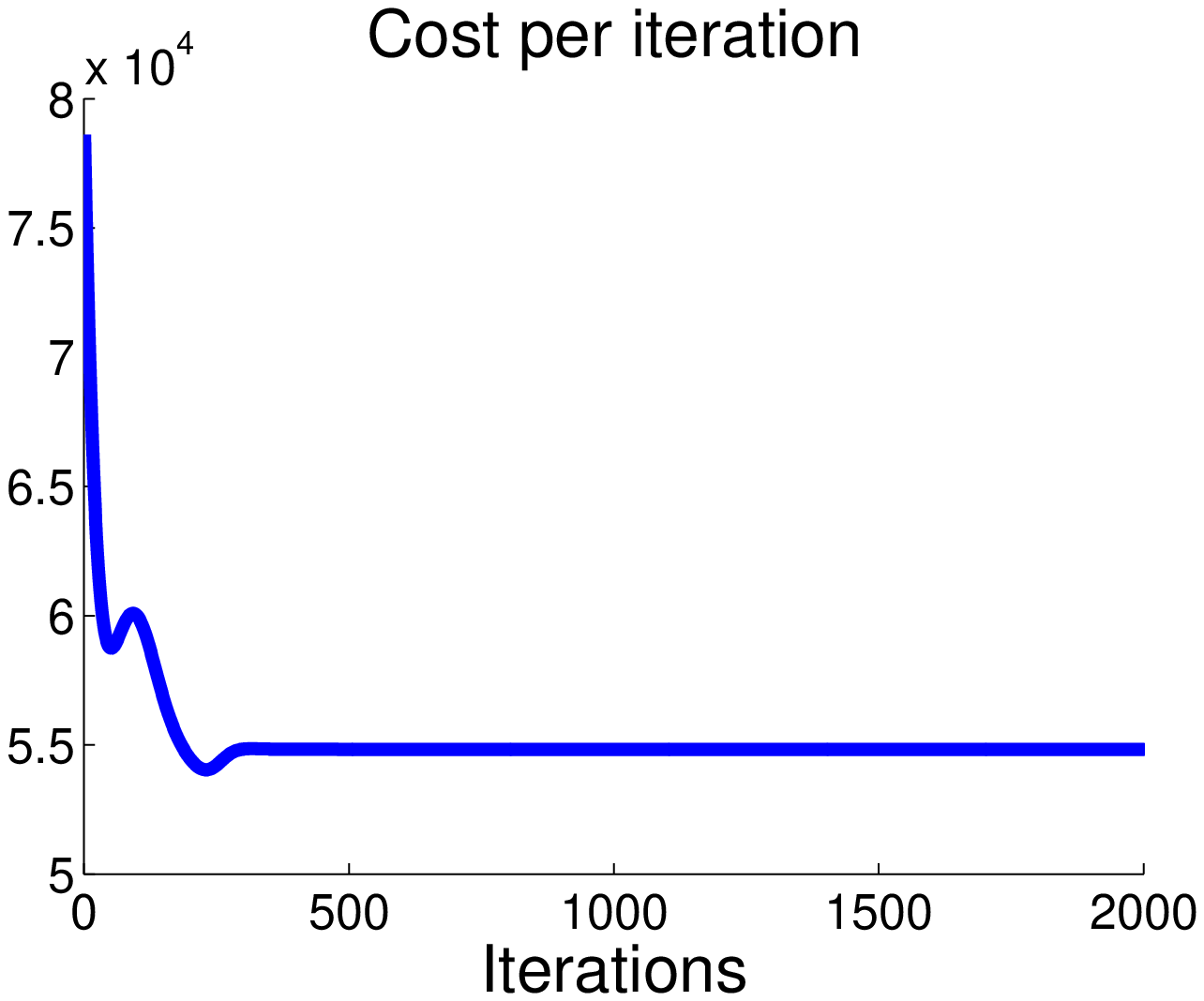}
                \caption{Cost per iteration.}
                \label{fig:quad_cost}
        \end{subfigure}%
        ~ %add desired spacing between images, e. g. ~, \quad, \qquad, \hfill etc.
          %(or a blank line to force the subfigure onto a new line)
        \begin{subfigure}[b]{0.3\textwidth}
                \includegraphics[width=\textwidth]{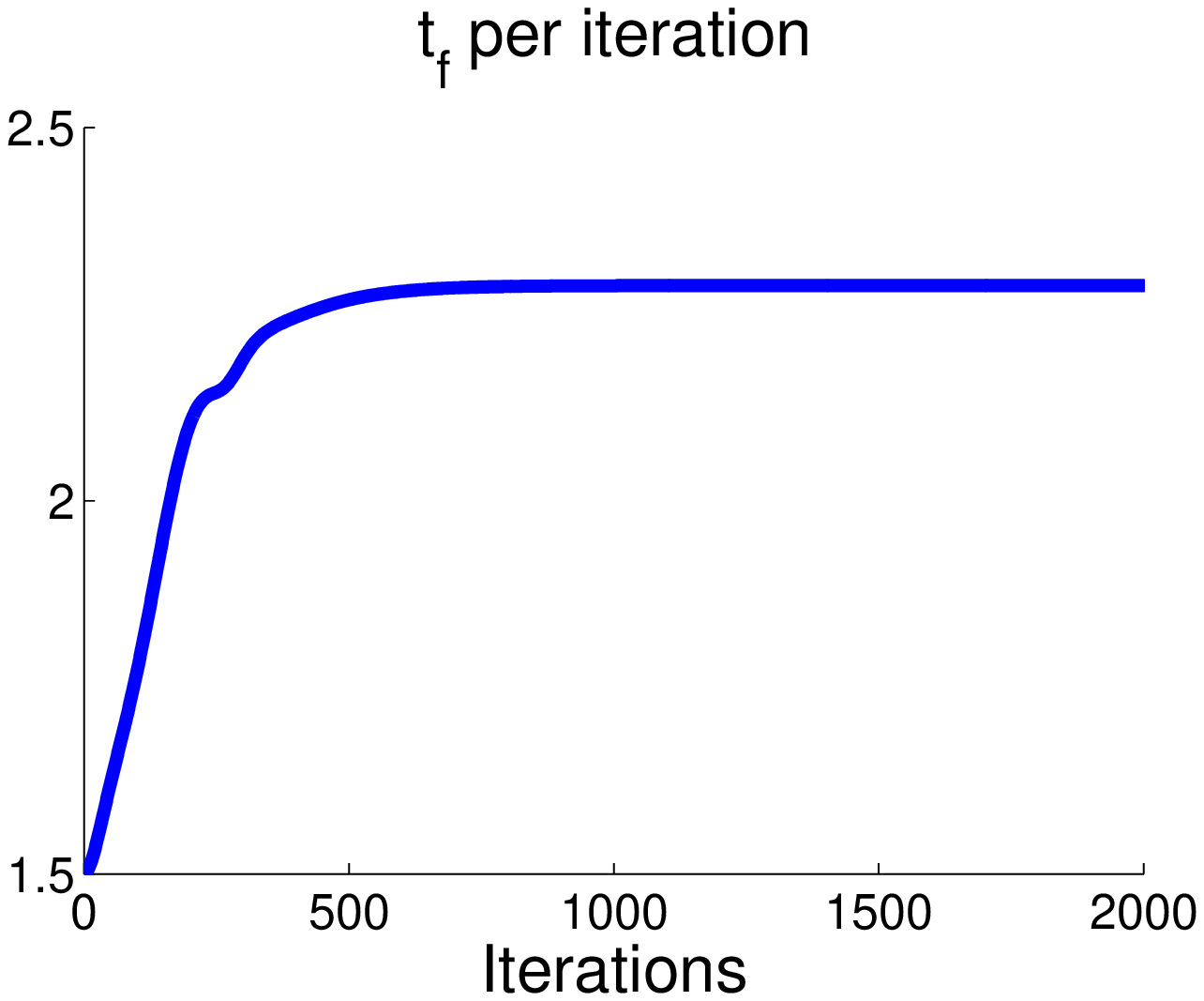}
                \caption{$t_f$ per iteration.}
                \label{fig:quad_tf}
        \end{subfigure}
        \caption{Cost and $t_f$ per iteration for the quadrotor.}\label{fig:quad_ctf}
\end{figure}

The dynamic model of the quadrotor includes 16 states: 3 for the position ($\vr = (x, y, z)^{\T}$), 3 for the Euler angles ($\Phi = (\phi, \theta, \psi)^{\T}$), 3 for the velocity ($\dot{\vr} = (\dot{x}, \dot{y}, \dot{z})^{\T}$), 3 for the body angular rates ($\dot{\Phi} = (p, q, r)^{\T}$) and 4 for the motor speeds ($\Omega = (\omega_1, \omega_2, \omega_3, \omega_4)^{\T}$). The corresponding dynamics of the quadrotor is given as follows:
\begin{align}
\frac{\rd \vx}{\rd t} = f(\vx) + G \vu,
\end{align}
where $\vx = [\vr, \Phi, \dot{\vr}, \dot{\Phi}, \Omega]^{\T} \in \Re^{16}$,  and $\vu = (u_1, u_2, u_3, u_4)^{\T} \in \Re^4$ is the control vector, where $u_1$ represents the thrust force, and $u_2, u_3, u_4$ represent the pitching, rolling, yawing moments, respectively.   The corresponding cost function is defined as $J = \frac{1}{2} \int_{0}^{\tf} [ c_t + (\vx - p)^{\T} Q (\vx - p) + \vu^{\T} R_{\vu} \vu ] + \frac{1}{2} (\vx(t_f) - p)^{\T} Q_f (\vx(t_f) - p) + \lambda^{\T} ([\vx_1 (t_f); \dots; \vx_6 (t_f)] - [p_1; \dots; p_6 ]),$ where $p = [p_1; \dots; p_{16}] \in \Re^{16}$ denotes the desired terminal states. In the simulation, we set 
\begin{align}
p(i) =  \begin{cases}
               1, i = 3;\\
               0, \text{otherwise},
            \end{cases} \quad \text{and} \quad
Q_f(i, i) =  \begin{cases}
               10^7, i = 1, 2, 3;\\
               10^6, i = 4, \dots, 9;\\
               10^5, i = 10, 11, 12;\\
               0, \text{otherwise},
            \end{cases}
\end{align}
and all the off-diagonal terms are assigned to $0$.  $Q = 0.01 Q_f.$ $R_{\vu} = 0.0001 I$. $\gamma = 0.02$ and $\epsilon = 0.02$. The desired terminal state $p$ is chosen for the quadrotor to execute the take-off maneuver. $50$ iterations are included to ensure the convergence and the cost per iteration is presented in Figure \ref{fig:quad_cost}. The corresponding optimal state trajectories are shown in Figure \ref{fig:optimal_quad_traj}. Optimal control $\vu$ is illustrated in Figure \ref{fig:optimal_quad_u}. $t_f$ per iteration is presented in Figure \ref{fig:quad_tf}.

              
\subsubsection*{Acknowledgments}

%\subsubsection*{References}

%\small{
%[1] Alexander, J.A. \& Mozer, M.C. (1995) Template-based algorithms
%for connectionist rule extraction. In G. Tesauro, D. S. Touretzky
%and T.K. Leen (eds.), {\it Advances in Neural Information Processing
%Systems 7}, pp. 609-616. Cambridge, MA: MIT Press.
%
%[2] Bower, J.M. \& Beeman, D. (1995) {\it The Book of GENESIS: Exploring
%Realistic Neural Models with the GEneral NEural SImulation System.}
%New York: TELOS/Springer-Verlag.
%
%[3] Hasselmo, M.E., Schnell, E. \& Barkai, E. (1995) Dynamics of learning
%and recall at excitatory recurrent synapses and cholinergic modulation
%in rat hippocampal region CA3. {\it Journal of Neuroscience}
%{\bf 15}(7):5249-5262.
%}

\bibliographystyle{unsrt}
\bibliography{references}

\end{document}